\documentclass[12pt]{article}

\newcommand{\C}[3]{\ifmmode C^{#1}_{#2#3}\else $C^{#1}_{#2#3}$\fi}
\newcommand{\Ch}[3]{\ifmmode \widehat C^{#1}_{#2#3}\else $\widehat 
C^{#1}_{#2#3}$\fi}
\newcommand{\D}[1]{\ensuremath{D(z_{#1})}}

\newcommand{\F}[3]{\ensuremath{{\cal F}_{#1#2#3}}}

\newcommand{\X}{\ifmmode X_{ijkm}\else $X_{ijkm}$\fi }
\newcommand{\Xh}{\ifmmode \widehat X_{ijkm}\else $\widehat X_{ijkm}$\fi}
\newcommand{\args}{\F11{},\dots,\F1\mathbin{{N-1}}{}}

\newcommand{\wb}[1]{\bar w_{#1}}

\newcommand{\dfrac}[2]{\displaystyle\frac{#1}{#2}}

\newcommand{\p}{\partial}
\newcommand{\pt}{\partial_{t_0}}

\newcommand{\sumN}[1]{\sum_{#1=1}^{N-1}}

\newcommand{\z}[1]{\ensuremath{z_{#1}}}
\renewcommand{\d}[3]{\ensuremath{D_{#1}D_{#2}D_{#3}}{\cal F}}
\newcommand{\dd}[2]{\ensuremath{D_{#1}D_{#2}{\cal F}}}

\renewcommand{\arraystretch}{1.3}

\catcode`\@=11
\def\marginnote#1{}

\newcount\hour
\newcount\minute
\newtoks\amorpm
\hour=\time\divide\hour by60
\minute=\time{\multiply\hour by60 \global\advance\minute by-\hour}
\edef\standardtime{{\ifnum\hour<12 \global\amorpm={am}%
           \else\global\amorpm={pm}\advance\hour by-12 \fi
           \ifnum\hour=0 \hour=12 \fi
           \number\hour:\ifnum\minute<10 0\fi\number\minute\the\amorpm}}
\edef\militarytime{\number\hour:\ifnum\minute<10 0\fi\number\minute}

\def\draftlabel#1{{\@bsphack\if@filesw {\let\thepage\relax
         \xdef\@gtempa{\write\@auxout{\string
             \newlabel{#1}{{\@currentlabel}{\thepage}}}}}\@gtempa
\if@nobreak
       \ifvmode\nobreak\fi\fi\fi\@esphack} \gdef\@eqnlabel{#1}}
\def\@eqnlabel{}
\def\@vacuum{}
\def\draftmarginnote#1{\marginpar{\raggedright\scriptsize\tt#1}}

\def\draft{
     \oddsidemargin -.5truein
     \def\@oddfoot{\footnotesize \sl preliminary draft: WDVV No. 16
\hfil
       \rm\thepage\hfil\sl\today\quad\militarytime}
     \let\@evenfoot\@oddfoot \overfullrule 3pt
       \let\label=\draftlabel
       \let\marginnote=\draftmarginnote
     \def\@eqnnum{(\theequation)\rlap{\kern\marginparsep\tt\@eqnlabel}%
       \global\let\@eqnlabel\@vacuum}

     }
\makeatletter
\newdimen\normalarrayskip              
\newdimen\minarrayskip                 
\normalarrayskip\baselineskip
\minarrayskip\jot
\newif\ifold             \oldtrue            \def\new{\oldfalse}
\def\arraymode{\ifold\relax\else\displaystyle\fi} 
\def\eqnumphantom{\phantom{(\theequation)}}     
\def\@arrayskip{\ifold\baselineskip\z@\lineskip\z@
        \else
        \baselineskip\minarrayskip\lineskip2\minarrayskip\fi}
\def\@arrayclassz{\ifcase \@lastchclass \@acolampacol \or
\@ampacol \or \or \or \@addamp \or
      \@acolampacol \or \@firstampfalse \@acol \fi
\edef\@preamble{\@preamble
     \ifcase \@chnum
        \hfil$\relax\arraymode\@sharp$\hfil
        \or $\relax\arraymode\@sharp$\hfil
        \or \hfil$\relax\arraymode\@sharp$\fi}}
\def\@array[#1]#2{\setbox\@arstrutbox=\hbox{\vrule
        height\arraystretch \ht\strutbox
        depth\arraystretch \dp\strutbox
        width\z@}\@mkpream{#2}\edef\@preamble{\halign
\noexpand\@halignto
\bgroup \tabskip\z@ \@arstrut \@preamble \tabskip\z@ \cr}%
\let\@startpbox\@@startpbox \let\@endpbox\@@endpbox
     \if #1t\vtop \else \if#1b\vbox \else \vcenter \fi\fi
     \bgroup \let\par\relax
     \let\@sharp##\let\protect\relax
     \@arrayskip\@preamble}
\def\eqnarray{\stepcounter{equation}%
                 \let\@currentlabel=\theequation
                 \global\@eqnswtrue
                 \global\@eqcnt\z@
                 \tabskip\@centering
                 \let\\=\@eqncr
    \halign to \displaywidth\bgroup
       \eqnumphantom\@eqnsel\hskip\@centering
       $\displaystyle \tabskip\z@ {##}$%
       \global\@eqcnt\@ne \hskip 2\arraycolsep
            $\displaystyle\arraymode{##}$\hfil
       \global\@eqcnt\tw@ \hskip 2\arraycolsep
            $\displaystyle\tabskip\z@{##}$\hfil
            \tabskip\@centering
       &{##}\tabskip\z@\cr}
\begingroup\ifx\undefined\newsymbol \else\def\input#1 {\endgroup}\fi
\newfont{\hr}{msbm10}
\newfont{\ams}{msam10}

\voffset= - 1.3in
\def\beq{\begin{equation}}
\def\eeq{\end{equation}}
\def\ba{\beq\new\begin{array}{c}}
\def\ea{\end{array}\eeq}
\def\be{\ba}
\def\ee{\ea}

\begin{document}
\begin{flushright}
FIAN/TD-06/01\\
ITEP/TH-22/01\\
SPIN-2001/10\\
EFI-2001-13\\
\end{flushright}
\vspace{0.2cm}
\begin{center}
  \textbf{\LARGE On Associativity Equations in
    Dispersionless Integrable Hierarchies}\\
\bigskip
\renewcommand{\thefootnote}{\alph{footnote}}
{
A.~Boyarsky}\\
{\em Spinoza Institute, Leuvenlaan 4, 3584 CE, Utrecht,
    the Netherlands
\footnote{On leave of absence from Bogoliubov ITP, Kiev,
    Ukraine; e-mail:\ A.M.Boyarsky@phys.uu.nl}},\\
\bigskip
{
A.~Marshakov}\\
{\em Theory
Department, Lebedev Physics Institute, Leninsky pr.~53, Moscow
~117924, Russia\\ and ITEP, Bol.Cheremushkinskaya str.~25,
Moscow~117259, Russia
\footnote{e-mail:\ mars@lpi.ru, andrei@heron.itep.ru}},\\
\bigskip
{
O.~Ruchayskiy}\\
{\em Enrico Fermi Institute and Department of Physics,
    The University of Chicago,\\
5640 S.Ellis Avenue, Chicago, IL 60637, USA
\footnote{e-mail:\ ruchay@flash.uchicago.edu}},\\
\bigskip
{
P.~Wiegmann}\\
{\em James Franck Institute and Enrico Fermi Institute,
    The University of Chicago,\\
5640 S.Ellis Avenue, Chicago,
    IL 60637, USA\\ and
    Landau Institute for Theoretical Physics
\footnote{e-mail:\ wiegmann@uchicago.edu}},\\
\bigskip
{
A.~Zabrodin}\\
{\em Institute of Biochemical Physics,
    Kosygina str.~4, Moscow~119991 GSP-1, Russia\\ and ITEP,
    Bol.~Cheremushkinskaya str.~25, Moscow~117259, Russia.
\footnote{e-mail:\ zabrodin@heron.itep.ru,
zabrodin@theory.uchicago.edu}}
\end{center} \setcounter{footnote}{0}

\begin{quotation}
  We discuss the origin of the associativity (WDVV) equations in the
context
  of quasiclassical or Whitham hierarchies. The associativity equations
are
  shown to be encoded in the dispersionless limit of the Hirota
equations for
  KP and Toda hierarchies.  We show, therefore, that any tau-function of
  dispersionless KP or Toda hierarchy provides a solution to
associativity
  equations.  In general, they depend on infinitely many variables.  We
also
  discuss the particular solution to the dispersionless Toda hierarchy
that
  describes conformal mappings and construct a family of new solutions
to the
  WDVV equations depending on finite number of variables.
\end{quotation}

\section{Introduction}
\label{sec:intro}

The Witten-Dijkgraaf-Verlinde-Verlinde (WDVV) equations \cite{WDVV} were
originally found in the context of 2D topological theories as an
associativity
condition of the operator algebra of primary fields. This is a highly
overdetermined system of nonlinear differential equations satisfied by
third-order derivatives of the ``free energy" function.  In particular,
solutions of the WDVV equations relevant to the simplest topological
Landau-Ginzburg models are of the form ${\cal F}=\log\tau$, where $\tau$
is
the dispersionless limit of the tau-function for a reductions of the KP
(Kadomtsev-Petviashvili) hierarchy of integrable equations \cite{LG}.
Thus one
may hope that a vast class of solutions can be constructed in a more
general
context of Whitham integrable hierarchies \cite{KriW}. Solutions to WDVV
equations were also found to be related to so-called Frobenius
manifolds~\cite{Dub}.

It turned out later that the associativity equations is a more general
phenomenon. For example, some new classes of solutions to the
associativity
equations were found~\cite{MMM} in the Seiberg-Witten theory, closely
related
to Whitham hierarchies (see also~\cite{Veselov,Luuk} for more recent
progress
in this direction), as well as in a more general context of integrable
hierarchies~\cite{Kri,vdLeur}. The origin of the WDVV equations in most
of
these cases is still a pending problem.

In this paper, we discuss a particular aspect of the WDVV
equations related to dispersionless integrable hierarchies.
Below we address the following two questions:
\begin{itemize}
\item Does any tau-function of the dispersionless KP (dKP)
   or Toda (dToda)
   hierarchies provide a solution to WDVV equations?
\item What are the conditions for these tau-functions
   (which are functions of an
   infinite number of variables) to obey a finite system of
WDVV equations?
\end{itemize}

Concerning the first question,
we show that {\em all} dispersionless tau-functions
do satisfy the WDVV equations or, in general, an infinite system of
equations of the WDVV type.
In other words, we show that the associativity equations can be
considered
as an intrinsic feature
of integrable hierarchies and furthermore
they can be derived from
the dispersionless limit of Hirota
relations by elementary manipulations.
 This result applies, in particular, to
the new class of dispersionless tau-functions, constructed
recently in connection with conformal maps~\cite{MWZWZ,KKMWZ}.
However, in generic situation the system of WDVV equations
obeyed by ${\cal F}= \log\tau$ is infinite, i.e.
both the number of equations and variables are infinite.
(In the language of topological theories
this means that the small phase space is not
  finite-dimensional).

Concerning the second question, we demonstrate the following:
A) Imposing a reduction of KP hierarchy
that leaves only finite number of
independent fields, it is
straightforward to show that the system of WDVV equations
becomes finite (a well known example is
the $N$-reduced
dKP hierarchy  relevant to topological
Landau-Ginzburg theories);
B) A new class of finite-dimensional solutions to WDVV
is obtained by a restriction of the tau-function for
conformal mappings~\cite{MWZWZ} to finite-dimensional
spaces of times related to conformal maps which are
Laurent polynomials of a fixed degree.

In Sec.~\ref{sec:notations} we introduce notations
and recall some standard
definitions from the literature
on WDVV equations. In Sec.~3 we show that any solution of the
dKP hierarchy obeys the infinite
system of WDVV equations. In Sec.~3.3 we obtain
the corresponding infinite-dimensional algebra
and the residue formula for third order derivatives
of the tau-function directly from the Hirota equation. In
Sec.~\ref{sec:n-kp} the reduction to their finite version is discussed.
Sec.~4.1 is devoted to the associativity equations in the
dToda hierarchy. In Sec.~4.2 we construct some
finite-dimensional solutions to the WDVV equations related to conformal maps.

\section{Associativity equations: the basic notation}
\label{sec:notations}

Let ${\cal F}$ be a differentiable function of variables $t_1 , t_2 ,
t_3,\ldots$ (``times''). Set
\begin{equation}
      \label{eq:2}
      \F ij{} \equiv \frac {\p^2 {\cal F}}{\p t_i \p t_j}\,,
\;\;\;\;
      \F ijk \equiv \frac {\p^3 {\cal F}}{\p t_i \p t_j \p t_k}
\end{equation}
for brevity. Let us choose one of these times, say $t_1$,
and assume that the matrix
$ \eta_{ij}=\F ij1$, called {\it metric},
is non-degenerate. Then one can
pass from the set of  variables $t_i$ to the set of variables ${\cal
F}_{j1}$. The matrix $C^l_{ij}=\p{\cal F}_{ij}/\p{\cal F}_{l1}$
connects
$ \F ijk$ and $ \eta_{ij}=\F ij1$:
\begin{equation}
\label{cf}
\F ijk{} =
\sum_l C_{ij}^l\F kl1.
\end{equation}
This equality can be understood as a definition of
$C_{ij}^{l}$:
$C_{ij}^{l}=\sum_k {\cal F}_{ijk}(\eta^{-1})^{kl}$, where
$\eta^{-1}$ is the inverse matrix to the $\eta$.

If one wants $C_{ij}^{l}$ to be structure
constants of an associative algebra
$\phi_i\cdot \phi_j = \sum_l C^k_{ij} \phi_k$,
the conditions
$\displaystyle{\sum_l C_{ij}^lC_{lk}^m = \sum_l C_{ik}^lC_{lj}^m}$
should be imposed. Equivalently
\be
\label{WDVV2}
\sum_l C_{ij}^l{\cal F}_{lkn} = \sum_l C_{ik}^l{\cal F}_{ljn}.
\ee
In other words
\beq\label{x}X_{ijkn}\equiv\sum_l C_{ij}^l{\cal F}_{lkn}
\eeq
is symmetric with respect to permutations of any indices.
The system~(\ref{WDVV2}) is called {\it associativity equations}
or {\it WDVV equations} (WDVV for short).

Assuming that eqs.\,(\ref{WDVV2}) hold,
one may choose any other index to define the metric $\eta (a)_{ij}
={\cal
F}_{a i j}$ (as long as it is non-degenerate) and structure
constants ${\cal F}_{ijk}=\sum_l C^{l}_{ij}(a)
{\cal F}_{kla}$.
They obey the same associativity relations~\cite{MMM}.
Introducing matrices
${\sf F}_{i}$ with matrix elements $({\sf F}_{i})_{jk}=
{\cal F}_{ijk}$, one may write the full set of WDVV equations
  in the form~\cite{MMM}:
\beq\label{fullWDVV}
{\sf F}_i {\sf F}^{-1}_{j}{\sf F}_{k}=
{\sf F}_k {\sf F}^{-1}_{j}{\sf F}_{i}
\;\;\;\;\;\;\mbox{for all}\;\; i,j,k.
\eeq

Eqs.\,(\ref{WDVV2}) with $C_{ij}^{l}$  defined
via~(\ref{cf}) are rather restrictive. They can be viewed as a
set of non-linear equations for the function ${\cal F}$ expressed
through its third order derivatives.
In the context of 2D topological theories and Seiberg-Witten theories
one is usually interested in solutions to WDVV with finite
number of variables,
i.e. the sum in~(\ref{WDVV2}) is finite\footnote{Besides,
two additional requirements
on solutions are usually imposed:
a) The metric is constant, i.e.
$\eta_{ij}$ does not depend on $t_k$; b) ${\cal F}$ is a
quasihomogeneous
function.
Sometimes these requirements are included~\cite{Dub}
into the definition of the WDVV system.}.
In this paper we
allow the number of variables to be infinite and
do not assume any special
properties of solutions. The only assumption is
that the series $\sum_{k}\frac{z^{-k}}{k}{\cal F}_{1k}$ defines
a function holomorphic and univalent in some domain which
includes infinity.

In the case of infinitely many variables
it is convenient to use generating functions for second and third order
derivatives of ${\cal F}$. Introduce the operator
\begin{equation}
      \label{eq:3}
      D(z) = \sum_{k=1}^\infty  \frac{z^{-k}}k \frac \p{\p t_k}.
\end{equation}
The  functions
\begin{equation}
     \label{eq:4}
     D_1D_2{\cal F} \equiv D(z_1)D(z_2){\cal F} = \sum_{k, m=1}^\infty
     \frac{\z1^{-k}}k \frac{\z2^{-m}}m \F{}km,
\end{equation}
\begin{equation}
     \label{eq:61}
     D_1D_2D_3{\cal F}
     \equiv D(\z1)D(\z2)D(\z3){\cal F} = \sum_{k, m, n=1}^\infty
     \frac{\z1^{-k}}k \frac{\z2^{-m}}m \frac {\z3^{-n}}n \F kmn
\end{equation}
generate the sets of $\F{}km$ and $\F kmn$. We also introduce
generating functions for the structure constants
\begin{equation}
      \label{eq:14}
       C^l (z_1 , z_2 )= \sum_{i, j=1}^\infty C^l_{ij}
\frac{\z1^{-i}}i \frac{\z2^{-j}}j,
\end{equation}
and for the $X_{ijkn}$~(\ref{x})
\begin{equation}
   \label{eq:16a}
   X(\z1,\z2,\z3,\z4) \equiv
\sum_{i,j,k,n=1}^{\infty}
     \frac{\z1^{-i}}{i} \frac{\z2^{-j}}{j} \frac {\z3^{-k}}{k}
\frac{z_{4}^{-n}}{n}
X_{ijkn}.
\end{equation}
The infinite WDVV equations~(\ref{WDVV2}) are then equivalent
to the symmetry of the $X(z_1, z_2 , z_3, z_4)$
under permutations of $z_1, z_2 , z_3 , z_4$.


\section{Hirota's relations for
dKP hierarchy and associativity equations}
\label{sec:Hirota-Identities-KP}
\subsection{The dKP hierarchy}\label{3.1}

Our starting point in this section is the bilinear identity for the
tau-function~\cite{Sato} which we refer to as Hirota equation.  Let
${\cal F}$
be the dispersionless limit of logarithm of the KP tau-function: ${\cal
  F}\equiv \log\tau$~\cite{KriW,TakTak}.  In the dispersionless limit
the
Hirota equation encodes a set of relations for the second order
derivatives
${\cal F}_{ij}$. In generating form they can be written as
\cite{TakTak,KodamaCarrol}:
\begin{equation}
      \label{eq:7}
      (\z1 - \z2) \left (1- e^{\D1\D2 {\cal F}}\right ) =
\left (\strut \D 1 - \D
        2\right )  \p_{t_1}{\cal F},
\end{equation}
where we use the operator~(\ref{eq:3}). The symmetric
version of this equation is
\begin{equation}
      \label{eq:8}
      (\z1-\z2)e^{\D1\D2 {\cal F}} +
      (\z2-\z3)e^{\D2\D3 {\cal F}} +
      (\z3-\z1)e^{\D3\D1 {\cal F}} = 0.
\end{equation}
Note that one can obtain~(\ref{eq:7}) from~(\ref{eq:8}) in the limit
$\z3
\rightarrow \infty$.
These equations should be understood as an infinite set
of algebraic relations for ${\cal F}_{ij}$ obtained
by expanding both sides as
a power series in $z_i$ and comparing the coefficients.
These relations can be
resolved with respect to ${\cal F}_{ij}$ with $i,j \geq 2$.
Indeed, writing~(\ref{eq:7}) as
\beq
\label{17}
D(z_1)D(z_2){\cal F}=\log \frac{p(z_1)-p(z_2)}{z_1-z_2}
\eeq
where
\beq
  \label{eq:10}
  p(z)=z-\sum_{k=1}^\infty\frac{z^{-k}}{k}{\cal  F}_{1k},
\eeq
we conclude that
$ {\cal F}_{ij}=P_{ij}({\cal F}_{11},\,
{\cal F}_{12},\, {\cal F}_{13}, \, \ldots \,) $,
with $P_{ij}$ being
  polynomials. This representation of the
dKP hierarchy goes back to~\cite{DB} (see
also~\cite{KodamaCarrol} for more details).

Second order derivatives of the tau-function allow one
to define a set of commuting
flows with generators $H_k$ determined from the series
\beq
\label{hams}
     D(z_1)D(z_2){\cal F}=-\log \left (1-\frac{z_2}{z_1}\right )-
\sum_{k=1}^{\infty}
     \frac{z_1^{-k}}{k}H_k(z_2).
\eeq
Acting by $D(z_3)$ on both sides and interchanging $z_1$
and $z_3$ one finds that
\beq
\label{commute}
\frac{\p H_i(z)}{\p t_j}=\frac{\p H_j(z)}{\p t_i}.
\eeq
Note that $H_1(z)=p(z)$~(\ref{eq:10}). The
relations~(\ref{commute}) can be viewed as a hierarchy
of evolution equations for the $p(z)$:
\beq\label{p}
\frac{\p p(z)}{\p t_k}=
\frac{\p H_k (z)}{\p t_1}
\eeq

We assume that the series $p(z)$ defines a function holomorphic
(except for a pole at infinity) and univalent in a domain including
infinity. Then the inverse function $z(p)$ is well defined as a
holomorphic univalent function in some domain around infinity in the
$p$-plane.
Equations~(\ref{p}), being rewritten as
evolution equations for the function $z(p)$, have the form of
dispersionless Lax-Sato equations:
\beq
     \label{Lax}
     \frac{\p z(p)}{\p{t_k}}=
\{H_k(p),\, z(p)\}_{KP},
\eeq
where the Poisson brackets are defined as
$\{f,g\}_{KP}=
\displaystyle{\frac{\p f}{\p p}\frac{\p g}{\p t_1}
-\frac{\p f}{\p t_1}\frac{\p g}{\p p}}$
and the derivatives in $t_i$ are taken at fixed $p$.
Moreover, as it follows from~(\ref{17}),
the $H_k$ turn out to be polynomials in $p$. On the other hand,
(\ref{hams}) fixes $H_k$ to be of the form
$H_k =z^k -D(z)\p_{t_k}{\cal F}$, i.e.
$H_k =z^k + O(z^{-1})$.
Therefore
\beq
\label{f}
H_k=(z^k(p))_{\geq 0},
\eeq
where the symbol $(f(p))_{\geq 0}$ means the non-negative part
of the Laurent series in $p$.
This is the dKP hierarchy (see
e.g.~\cite{TakTak}).
Given a Lax function $z(p)=p+O(z^{-1})$ and $H_k$ obtained
from it by means of~(\ref{f}), one can reconstruct the
second order derivatives ${\cal F}_{jk}$ via the formula
\beq\label{res2ord}
{\cal F}_{jk}=\frac{1}{j+k}\,\mbox{res}_{\infty}
\left ( \frac{dH_j dH_k}{d \log z}\right ).
\eeq

A remark is in order.  Note that eqs.~(\ref{hams}--\ref{Lax}) hold for
any
function ${\cal F}$.  We want to stress that this is not an integrable
hierarchy yet, in spite of the fact that there are infinitely many
commuting
flows~(\ref{commute}). The crucial relation, which really makes an
integrable
hierarchy out of this, is~(\ref{f}). The Hirota equation gives a
relation
between the generating function of the flows and $p$ and allows one to
determine $H_k$ as functions $H_k(p)$ with certain analytic properties.
In the
dKP-case they are polynomials. From this point of view, it is the Hirota
identity that encodes integrability of the system.

Plugging~(\ref{17}) in the r.h.s. of~(\ref{hams}) and differentiating
w.r.t.
$p=p(z_2)$, one arrives at the relation (cf.~\cite{KodamaCarrol})
\begin{equation}
\label{gener}
\frac{1}{p(z_1)-p}=\sum_{k\geq 1} \frac{z_{1}^{-k}}{k}\,
\frac{d H_k (p)}{d p}.
\end{equation}
It is used below to obtain an explicit realization of the associative
algebra.

Let us stress that all basic relations of the dKP hierarchy discussed in
this
section contain second order derivatives of ${\cal F}$ only.  We are
going to
take an extra derivative and rearrange the resulting relations into the
form
of the associativity equations.

\subsection{Proof of WDVV equations}
\label{sec:WDVV-from-Hirota}

An elementary manipulation with the Hirota equation~(\ref{eq:7}) allows
one to
bring it to the form~(\ref{cf}), which is the defining relation for the
structure constants $C_{ij}^{l}$.  Apply $D(z_3)$ to the both sides
of~(\ref{eq:7}). This gives
\begin{equation}
      \label{eq:11}
D_1 D_2 D_3 {\cal F} = -\, \frac{1}{p_1 - p_2}(D_1 -
        D_2)D_3\p_{t_1}{\cal F},
\end{equation}
where $p_i \equiv p(z_i)$.

We observe that  eq.~(\ref{eq:11}), being written in modes
(by means of~(\ref{eq:3},~\ref{eq:61},~\ref{eq:14}))
is equivalent to
the infinite-dimensional version of~(\ref{cf}):
\begin{equation}
      \label{eq:15}
      \F ijk = \sum_{l=1}^\infty C^l_{ij} \F lk1
\end{equation}
where
the structure constants are defined by the
generating function
\begin{equation}
      \label{eq:142}
      C^l(z_1 , z_2 )=
      -\, \frac{\z1^{-l} - \z2^{-l}}{l(p_1 - p_2)}.
\end{equation}
Since ${\cal F}$ obeys~(\ref{eq:7}), one may rewrite
it in an equivalent form:
\begin{equation}
      \label{eq:142a}
      C^l(z_1 , z_2 )=
      - \, \frac{\z1^{-l} - \z2^{-l}}{l(z_1 - z_2)}
\, e^{-D(z_1)D(z_2) {\cal F}}.
\end{equation}
It is easy to see from~(\ref{eq:14}) that the infinite
sum in~(\ref{eq:15}) is actually always finite: it truncates
at $l=i+j$.

Let us show that
${\cal F}$ obeys the
WDVV~(\ref{WDVV2}), with each index running
over natural numbers.
In terms of generating functions this means that
$X(z_1, z_2 , z_3 , z_4 )$ given by~(\ref{eq:16a})
is totally symmetric w.r.t. permutations of \z1\dots\z4.
It is enough to prove the symmetry w.r.t. the permutation
of $z_2$ and $z_3$, which is equivalent to the relation
\begin{equation}
   \label{eq:18}
   z_{13}e^{D_1D_3{\cal F}}(D_1-D_2)D_3D_4{\cal F} =
z_{12}e^{D_1D_2{\cal
       F}}(D_1-D_3)D_2D_4{\cal F},
\end{equation}
where $z_{ik}=z_i - z_k$.
Using~(\ref{eq:8}) it is straightforward to bring~(\ref{eq:18})
into the form
\begin{equation}
\label{eq:22}
D_4 \biggl(z_{13}e^{D_1D_3{\cal F}} - z_{12}e^{D_1D_2{\cal
     F}} - z_{23}e^{D_2D_3{\cal F}} \biggr) = 0
\end{equation}
which is the $D_4$-derivative of~(\ref{eq:8}) and therefore
the WDVV follows from the Hirota equation.

One can prove in a similar way WDVV for ${\cal F}$
choosing $D(z_a){\cal F}_{ij}$ rather than ${\cal F}_{1ij}$
as a metric.
Apply $D(z_3)$ to the symmetric form of the Hirota equation
(\ref{eq:8}) written for the three points $z_1 , z_2 , z_a$.
One obtains:
\be
\label{sym}
D_1D_2D_3{\cal F} = \frac{1}{p_{12}}
(p_{1a}D_1 - {p_{2a}}D_2)D_3 D_a {\cal F},
\ee
where
$p_{ik}\equiv (\z i - \z k)e^{D_i D_k {\cal F}}$.
Similarly to the previous case, which is reproduced in the
limit $z_a \to \infty$, this equality defines structure
constants.
The WDVV equations are then equivalent to
\be
\label{eqg}
p_{1a}\left(p_{13}D_1D_3D_4{\cal F} - p_{12}D_1D_2D_4{\cal F}\right) =
\left(p_{13}p_{2a} - p_{12}p_{3a}\right)D_2D_3D_4{\cal F}
\ee
Plugging the $D_4$-derivative of~(\ref{eq:8}),
write the l.h.s. of~(\ref{eqg})  as
$p_{1a}p_{23}D_2D_3D_4{\cal F}$. It is clear then that~(\ref{eqg}) is
equivalent to the identity
$p_{1a}p_{23} = p_{13}p_{2a} - p_{12}p_{3a}$
which is automatically satisfied by
$p_{ij} = p_i - p_j$, as is indeed the case
due to~(\ref{eq:8}).

\subsection{Realization of associative algebra and the residue formula}

To give a realization of the associative algebra with the
structure constants defined by~(\ref{eq:142}), we introduce
the polynomials
\beq\label{basis}
\phi_k (p) =\frac{d H_k (p)}{d p}\,,
\;\;\;\;\;\;k\geq 1.
\eeq
Expanding both sides of the identity
$\displaystyle{\frac{1}{(p-p_1)(p-p_2)}=\frac{1}{p_1 -p_2}\left(
\frac{1}{p-p_1}-\frac{1}{p-p_2}\right)}$ in $z_1^{-1}$, $z_2^{-1}$,
using
(\ref{gener}), and comparing the coefficients,
we obtain the algebra
\beq\label{alg1}
\phi_i (p) \phi_j (p) =\sum_{l\geq 1}C_{ij}^{l}\phi_l (p)
\eeq
where the structure constants are exactly those defined
by~(\ref{eq:142}). This infinite-dimensional
algebra is just the ring of polynomials of arbitrary degree.

For completeness, we show how to derive the
residue formula~\cite{KriW,A-K}
for third order derivatives of ${\cal F}$ directly
from the Hirota equation. Substituting into
(\ref{eq:11}) its particular case $D_1D_2\p_{t_1}{\cal F} =
\displaystyle{-\frac{(D_1-D_2)
\p_{t_1}^2{\cal F}}{p_1 -p_2}}$ (obtained in the
limit $z_3\to\infty$),
one easily expresses
$D_1 D_2 D_3 {\cal F}$ in terms of $D_i \p^{2}_{t_1}{\cal F}$
only:
$$
D_1 D_2 D_3 {\cal F}=\sum_{i =1}^{3}\mbox{res}_{p_i}
\left ( \frac{D(z(p)) \p^{2}_{t_1}{\cal F}}{(p-p_1)(p-p_2)(p-p_3)}dp
\right ).
$$
In the numerator we have: $D(z)\p^{2}_{t_1}{\cal F}=
-\p p(z)/\p t_1$ which is equal to $\p_{t_1}z(p)/z'(p)$ in terms of
the independent variable $p$ ($z'(p)\equiv dz/dp$). Expanding both sides
of the above formula in the series in $z_{1,2,3}$ and using
(\ref{gener}) we obtain:
\beq\label{res1}
{\cal F}_{jkm}=\frac{1}{2\pi i}\oint_{C_{\infty}}\!
\frac{\p_{t_1}z(p)}{z'(p)}
\phi_j (p) \phi_k (p) \phi_m (p) dp,
\eeq
where $C_{\infty}$ is a small contour around infinity in the domain
where
$p(z)$ is holomorphic and univalent. It suffices that $z'(p)$
does not have
zeros and singularities in the domain. One may think of~(\ref{res1})
as coming from the scalar product
\beq\label{res2}
\langle f\cdot g\rangle =\frac{1}{2\pi i}\oint_{C_{\infty}}\!
\frac{\p_{t_1}z(p)}{z'(p)}
f(p)g(p) dp.
\eeq
defined for polynomials.
Note that $\eta_{jk}={\cal F}_{jk1}=
\langle \phi_j \cdot \phi_k \rangle$ (just because $\phi_1 =1$).
Therefore, the algebra~(\ref{alg1}) is in full agreement
with~(\ref{cf}):
\beq\label{proof}
{\cal F}_{jkm}=\langle \phi_j \phi_k \cdot \phi_m \rangle
=\sum_{l}C_{jk}^{l}
\langle \phi_l \cdot \phi_m \rangle
=\sum_{l}C_{jk}^{l}{\cal F}_{lm1}.
\eeq
Meanwhile, this gives another proof of WDVV for ${\cal F}$:
$X_{jkmn} \equiv\langle
\phi_j \phi_k  \cdot \phi_m  \phi_n \rangle$ (this
representation is obviously
equivalent to~(\ref{x}))
is symmetric due to apparent symmetry of the r.h.s. of~(\ref{res1}).

If there exists a times-independent function $\varphi (z)$ such that
$E(p)\equiv\varphi (z(p))$ is a meromorphic function of $p$
with the number of poles being unchanged under variations of all
$t_j$, then the integral is equal to the sum of
residues at zeros of $E'(p)$ (cf.~\cite{A-K}):
\beq\label{res3}
{\cal F}_{jkm}=
\sum_{E'(p_{a})=0} \mbox{res}_{p_a} \left (
\frac{\p_{t_1}E(p)}{E'(p)}
\phi_j (p) \phi_k (p) \phi_m (p) dp \right ).
\eeq
Indeed, $\p_{t_1}z(p)/z'(p) =\p_{t_1}E(p)/E'(p)$
and poles of $\p_{t_1}E(p)$
do not contribute to the integral since they are canceled
by those of $E'(p)$.

The existence of such a function $E$ means a finite-dimensional
reduction of the hierarchy. In this case the WDVV algebra
becomes finite-dimensional. This can be easily seen directly
from the residue formula.
Below we show this starting from the Hirota equation.

\subsection{Finite-dimensional reductions} \label{sec:n-kp}

In terms of the Hirota equation the finite-dimensional reduction
is a set of additional constraints for
second order derivatives of tau-function:
\begin{equation}
      \label{eq:28}
      \F{}1M = Q_M (\args)\,,
\;\;\;\;\;\; M\geq N
\end{equation}
where functions $Q_M$ do not explicitly depend on times and are
required to be consistent with the evolution equations.
Any reduction of this kind leaves us with
$N-1$ independent primary fields
which can be chosen to be ${\cal F}_{1j}$ with $1\leq j \leq N-1$.
All other ${\cal F}_{1M}$ with $M\geq N$ (descendants) are
expressed through the independent ones via formulas~(\ref{eq:28}).

A particular example is the familiar $N$-KdV
reduction, for which ${\cal F}$ is
independent of $t_{N},\,t_{2N},\,t_{3N},\dots$
In this case functions $Q_M$ are certain
polynomials with rational coefficients such that $E(p)=z^N(p)$
is a polynomial. Other reductions,
when $z^N$ is a rational function, are also known~\cite{Kr1,A-K}.
In the sequel we do not refer to explicit form of~(\ref{eq:28}).

Hereafter, we use capital letters (like $J,M,L, \ldots$) for
the descendants, i.e. for indices larger than $N-1$.
     Summation over {\it small} (primary) indices
(from $1$ to $N-1$) is always indicated explicitly.

In the case of a reduction the metric
${\cal F}_{lk1}$ in~(\ref{eq:15}) becomes a
degenerate matrix of  rank $N-1$.
Indeed, taking the derivative of~(\ref{eq:28})
w.r.t. $t_k$, we find
that $J$-th line
of the matrix ${\cal F}_{jk1}$
is a linear combination of the first $N-1$ lines.
The finite WDVV equations are obtained by means of a
projection on the nondegenerate
$N-1$-dimensional subspace.

     Let us separate the
sums over primary and descendant fields in~(\ref{eq:15}) and use
(\ref{eq:28}) to express ${\cal F}_{Lk1}$ through ${\cal F}_{lk1}$:
$$
\begin{array}{lll}
{\cal F}_{ijk}
=\displaystyle{
\sum_{l=1}^{N-1}\left (
        C_{ij}^{l} +
\sum_{L} C_{ij}^{L}\,\frac{\p Q_L}{\p {\cal F}_{1l}}
\right )
{\cal F}_{lk1}}.
\end{array}
$$
The object
\begin{equation}
      \label{eq:32}
      \tilde C_{ij}^{l}
=C_{ij}^{l} +
\sum_{L} C_{ij}^{L}\,\frac{\p Q_L}{\p {\cal F}_{1l}},
\end{equation}
at $i,j<N$
defines the structure constants of a finite dimensional
algebra formed by the primary fields.
The system~(\ref{eq:15}) becomes finite:
\begin{equation}
      \label{eq:33}
      \F ijk = \sumN{l}
      \tilde C_{ij}^{l}
     \F lk1,
\end{equation}
where the metric $\F lk1$ is non-degenerate
on the small space ($l,\,k=1,\dots, N-1$).
The
structure constants of the primary fields obey the
  finite-dimensional WDVV equations. In other words,
$\tilde X_{ijkm}$ defined by
      $\displaystyle{ \tilde X_{ijkm}  \equiv \sumN{l}
      \tilde C_{ij}^{l} \F lkm }$
is  symmetric with respect to the permutations of
the (small) indices $i,j,k,m$.
This follows from the fact that
$\tilde X_{ijkm} = \X$ for $i,j,k,m<N$.

The proof is elementary. We write
\begin{equation}
      \label{eq:40}
      \X = \sumN{l} \C lij \F lkm + \sum_{L} \C Lij \F Lkm,
\end{equation}
and substitute $\C lij =
      \tilde C_{ij}^{l}
-\sum_{L} \C Lij \, \p Q_L /\p {\cal F}_{1l}$
into
eq.~(\ref{eq:40}):
     $\displaystyle{ X_{ijkm} = \tilde X_{ijkm}
      + \sum_{L} \C Lij Y_{Lkm}}$,
where
$Y_{Lkm}= \displaystyle{
{\cal F}_{Lkm}-
\sum_{l<N} \frac{\p Q_L}{\p {\cal F}_{1l}} \F lkm }$
vanishes.
To see this, express ${\cal F}_{lkm}$ inside the sum in terms of
${\cal F}_{ij1}$ using~(\ref{eq:33})
and interchange the order of summation.
Using~(\ref{eq:33}) once again, one obtains the result.

\section{Associativity equations in dToda
hierarchy}

\subsection{Hirota relations for dToda hierarchy
and WDVV equations}
\label{sec:toda}

The dToda
hierarchy reveals the WDVV algebra in a similar manner.
The arguments almost completely repeat those of the
previous sections, the Hirota equations being somewhat
different.
The independent
variables of the dToda hierarchy  are
$t_k$, where
$k$ is any integer number:
$t_0,t_{\pm 1},t_{\pm 2},\dots$
Let us introduce two functions:
\begin{equation}
      \label{eq:48}
       w^{\pm}(z) = z \exp \left (-\frac 12 \p_{t_0}^2{\cal F}
-\p_{t_0}D^{\pm}(z){\cal F} \right )
\end{equation}
where $D^{\pm}( z)=\displaystyle{\sum_{k =1}^{\infty}\frac{z^{-k}}{k}\,
\frac{\p}{\p t_{\pm k}}}$
and assume that they are regular and univalent in
some domain around infinity.
The Hirota equations for the
dToda hierarchy read~\cite{TakTak,KKMWZ}
\begin{equation}
      \label{eq:50}
      w^{\pm}(z_1)-w^{\pm}(z_2) =
      (\z1 -\z2)\,e^{-\frac 12 \p_{t_0}^2{\cal F}} e^{-D^{\pm}_{1}
D^{\pm}_{2}{\cal F }},
\end{equation}
\begin{equation}
      \label{eq:64}
1 - \frac 1 {w^+ (z_1) w^- (z_2)}= e^{-D^+_1 D^-_2 {\cal F}}.
\end{equation}

Similarly to the arguments of
Sec.~\ref{3.1} Hirota equations~(\ref{eq:50},~\ref{eq:64}) define the
dToda hierarchy with commuting flows generated by
\beq
\label{H}
H_{\pm j}(w)=\Bigl ((z^\pm(w^{\pm 1}))^j\Bigr )_{\pm} +\frac{1}{2}
\Bigl ((z^\pm(w^{\pm 1}))^j \Bigr )_{0},\quad j\geq 1\,,
\;\;\;\;\;
H_0(w)=\log w.
\eeq
Here $z^{\pm}(w)$ is the inverse function of $w^{\pm}(z)$
and $(...)_{\pm}$, $(...)_{0}$ means strictly positive, negative,
and constant part of the Laurent series, respectively.
The Lax-Sato equations read
\beq\label{T1}
\frac{\p z^{\pm}(w^{\pm 1})}{\p t_j}=\mbox{sign}\,j \,
\{ H_j (w) ,\, z^{\pm}(w^{\pm 1})\}_{Toda},
\eeq
where the Poisson bracket is defined as
$\{f,g\}_{Toda}=\displaystyle{
w\frac{\p f}{\p w} \frac{\p g}{\p t_0}
- w\frac{\p f}{\p t_0} \frac{\p g}{\p w}}\,$.
The dToda analog of the generating function~(\ref{gener})
for derivatives of $H_j$ is
\beq\label{gener1}
\frac{w^{\pm 1}}{w^{\pm}(z_1)-w^{\pm 1}}=
\pm \sum_{k\geq 1} \frac{z_{1}^{-k}}{k}\,
w\frac{d  }{d w}H_{\pm k}(w).
\eeq

We define the metric to be
$\eta_{ij}={\cal F}_{0ij}$,
where the indices take integer values,
and the structure constants as
\beq\label{stct}
{\cal F}_{ijk}=\sum_{l=-\infty}^{\infty}
C_{ij}^{l}{\cal F}_{lk0}.
\eeq
 From the definition of the metric, we have
$C_{0j}^{l}=\delta^{l}_{j}$ for all $l,j$.
To find other structure constants, apply $\p_{t_k}$
to the Hirota equations~(\ref{eq:50}) and~(\ref{eq:64}):
\begin{equation}
      \label{eq:51}
      D^{\pm}_{1} D^{\pm}_{2}
\p_{t_k} {\cal F}=
\frac{1}{w^{\pm}_{1} - w^{\pm}_{2} }\,
\left (
w^{\pm}_{2} D^{\pm}_{2} {\cal F}_{k0}
- w^{\pm}_{1} D^{\pm}_{1} {\cal F}_{k0}
\right ),
\end{equation}
\begin{equation}
\label{eq:65}
D^{+}_{1} D^{-}_{2}
\p_{t_k} {\cal F}=
\frac{1}{w^{+}_{1} w^{-}_{2} -1 }\,
\left (
{\cal F}_{k00} +
D^{+}_{1} {\cal F}_{k0}+
D^{-}_{2} {\cal F}_{k0}
\right ).
\end{equation}
Here and below $w^{\pm}_i =w^{\pm}(z_i)$.
In complete analogy with eq.~(\ref{eq:14})
generating functions for structure constants are read from
the r.h.s. of these equations. We conclude  from~(\ref{eq:51}) that
$C_{ij}^{l}=0$ whenever $i,j$ are both positive and
$l\leq 0$
or both negative and $l\geq 0$.
If all the indices are
positive or all negative, we have:
\begin{equation}
      \label{eq:56}
       \sum_{\pm i \geq 1}
       \sum_{\pm j \geq 1}
\C l ij
\frac{z_{1}^{\mp i}}{i}
\frac{z_{2}^{\mp j}}{j}
=-\,\frac{
w^{\pm}_{1}z_{1}^{\mp l}-
w^{\pm}_{2}z_{2}^{\mp l}}{\pm l (
w^{\pm}_{1} -w^{\pm}_{2})}\,,
\qquad
\pm l \geq 1.
\end{equation}
When $i$ and $j$ have different signs we use
eq.\,(\ref{eq:65}) to obtain:
\begin{equation}
      \label{eq:69}
       \sum_{i \geq 1}
       \sum_{j \leq -1}
\C l ij
\frac{z_{1}^{-i}}{i}
\frac{z_{2}^{j}}{j}
=\frac{z_{1}^{-l}}{l (
1- w^{+}_{1}w^{-}_{2})}\,,
\qquad  l \geq 1,
\end{equation}
\begin{equation}
      \label{eq:69a}
       \sum_{i \geq 1}
       \sum_{j \leq -1}
\C l ij
\frac{z_{1}^{-i}}{i}
\frac{z_{2}^{j}}{j}
=-\, \frac{z_{2}^{l}}{l (
1- w^{+}_{1}w^{-}_{2})}\,,
\qquad l \leq -1,
\end{equation}
\begin{equation}
      \label{eq:69b}
       \sum_{i \geq 1}
       \sum_{j \leq -1}
C^{0}_{ij}
\frac{z_{1}^{-i}}{i}
\frac{z_{2}^{j}}{j}
=\frac{1}{1- w^{+}_{1}w^{-}_{2} }.
\end{equation}
With this definition of the structure constants
at hand, one can prove WDVV for
any solution to the dToda hierarchy\footnote{In a
very particular case, the infinite WDVV in dToda hierarchy
were found in~\cite{BonoraXiong}.}
\begin{equation}\label{WDVVToda}
  \sum_{l=-\infty}^{\infty}
  C_{ij}^{l}{\cal F}_{lkm}=
  \sum_{l=-\infty}^{\infty}
  C_{ik}^{l}{\cal F}_{ljm}
\end{equation}
in the same way as for the dKP-case.
The calculation is somewhat
longer since the cases when all indices have
different possible signs have to be considered
separately. Details of the proof are given in the
appendix.

The realization of the associative algebra defined by the structure
constants~(\ref{eq:56})--(\ref{eq:69b}) is obtained with the help
of eq.\,(\ref{gener1}) exactly in the same way as in the dKP-case.
The  generators
\beq
\label{infring}
\phi_i(w) = w{dH_i\over dw}
\eeq
for all integer $i$
span the ring of Laurent polynomials of arbitrary degree. In this basis
the structure constants of the algebra are given by
(\ref{eq:69}-\ref{eq:69b}).

The derivation of the residue formulas from
Hirota relations is also parallel to the dKP-case.
Consider for simplicity the case when all the indices
are positive. We have:
$$
D^+_1 D^+_2 D^+_3 {\cal F}=\sum_{\alpha =1}^{3}\mbox{res}_{w_{\alpha}}
\left ( \frac{D^+(z^+(w)) \p^{2}_{t_1}{\cal F}}{(w-w_1)(w-w_2)(w-w_3)}dw
\right ).
$$
By virtue of~(\ref{gener1}),
this is equivalent to
\beq\label{res11}
{\cal F}_{jkm}=\frac{1}{2\pi i}\oint_{C_{\infty}}\!
\frac{\p_{t_0}z^+(w)}{  z^{+}(w)'}\,
\phi_j (w) \phi_k (w) \phi_m (w) \frac{dw}{w^2},\quad
j,k,m \geq 1,
\eeq
where again $z^+(w)' = dz^+ /dw$.
Similar formulas can be written for non-positive indices.


\subsection{The associative algebra related to conformal maps}
\label{sec:assoc-algebra-relat}

As shown in~\cite{MWZWZ},
a particular solution ${\cal F}^{(0)}$ to the dToda hierarchy
describes evolution of
conformal mapping of a complex domain with respect to
deformations of the domain.
This solution is specified by the reality
conditions
$t_{-k}=\bar t_k$, $w^{-}(z)=\bar w^+(z)$\footnote{Here
and below bar means
complex conjugation and for any series $f(z)=\sum f_k z^k$
we set $\bar f (z)=\sum \bar f_k z^k$.}
consistent with the hierarchy. Under the reality
conditions ${\cal F}^{(0)}$ is a real-valued function of times.
The relation to conformal maps
is as follows. Let
$\displaystyle{z(w)=rw+\sum_{k\geq 0}u_k z^{-k}}$ be the
univalent conformal map\footnote{To ensure conformality and
univalentness, absolute values of the
coefficients $u_k$ should be restricted by some inequalities
which we do not discuss here (see e.g.~\cite{Helle}).}
from the exterior of the unit circle $|w|>1$ to the exterior
of a given analytic curve $\gamma$, the normalization being fixed
by the conditions that infinity is taken to infinity and
$r$ is real and positive.
Then we set $z^+(w)=z(w)$, $z^-(w)=\bar z(w^{-1})$. The function
$w(z)=w^+(z)$ is the inverse map.
It has been shown~\cite{MWZWZ}
that evolution of the map is described by the dToda hierarchy with the
generators of commuting flows given by~(\ref{H}). The reality conditions
imply $H_{-j}(w)=\bar H_{j}(w^{-1})$.  The times are harmonic moments of
the
exterior domain:
\beq t_k=\frac{1}{2\pi i k}\oint_\gamma z^{-k}\bar zdz
\eeq
with the origin assumed to be outside the domain.
The ``initial conditions'' for the
solution are given
by the dispersionless limit of the string equation:
\beq\label{string}
\{z(w),\, \bar z(w^{-1})\}_{Toda}=1.
\eeq

In this setting the residue formula~(\ref{res11})
can be written in a more transparent form.
Since $z(w)$ maps the exterior of the unit circle in a conformal manner,
for some region in the space of $t_k$,
neither zeros of $z'(w)$ nor poles or other singularities
of $z(w)$ are in the domain $|w|>1$.
Therefore, the function under the integral in~(\ref{res11})
is regular everywhere outside the unit circle except infinity.
So, the integration contour can be taken to be the
unit circle $|w|=1$:
\beq\label{res111}
{\cal F}_{jkm}^{(0)}=\frac{1}{2\pi i}\oint_{|w|=1}\!
\frac{\p_{t_0}z(w)}{  z'(w)}\,
\phi_j (w) \phi_k (w) \phi_m (w) \frac{dw}{w^2},\quad
j,k,m \geq 1.
\eeq
The string equation~(\ref{string}) reads
$$
\frac{\p_{t_0}z(w)}{z'(w)}=
- w^2\, \frac{\p_{t_0}\bar z(w^{-1})}{\bar z'(w^{-1})}+ \,
\frac{w}{z'(w) \bar z'(w^{-1})}
$$
where $\bar z'(w^{-1})$ is the derivative $d\bar z/dw$ taken at the
point $w^{-1}$.
Plugging this into~(\ref{res111})
and taking into account that
the function $\p_{t_0}\bar z (w^{-1})/\bar z'(w^{-1})$
is regular inside the unit circle, we come to
\beq\label{reskri}
{\cal F}_{jkm}^{(0)} =
\,\frac{1}{2\pi i}\oint_{|w|=1}
\frac{\phi_j(w) \phi_k(w) \phi_m (w)}{z'(w) \bar z'(w^{-1})}
\,{dw\over w}\
= -\,\frac{1}{2\pi i}\oint_{\gamma}
\frac{dH_j dH_k dH_m}{dz d\bar z}.
\eeq
A more detailed analysis shows that this formula is valid,
up to an overall sign,
for all integer indices, not only for positive ones.
This formula
was first derived by I.Krichever~\cite{Kunpublished}
within a different approach.

One may again interpret~(\ref{reskri})
as the scalar product
\beq
\label{scalar}
\langle f\cdot g\rangle= \,\frac{1}{2\pi i}\oint_{|w|=1}
\frac{dw} {wz'(w) \bar z'(w^{-1})}\ f(w)g(w)
\eeq
on the infinite-dimensional ring of
Laurent polynomials. In the basis~(\ref{infring}) the
algebra reads
$\displaystyle{\phi_i(w)\phi_j(w) = \sum_k C_{ij}^k \phi_k(w)}$,
where the structure constants
are given by
(\ref{eq:56})-(\ref{eq:69b}). The proof is the same as for
(\ref{proof}).

Comparing with the dKP residue formula~(\ref{res3}), one may say that
the
requirement of reality together with that of conformality and
univalentness
effectively defines a reduction: in both cases these conditions ensure
that the integral in~(\ref{res11}) is saturated by singularities coming
from
the denominator. From this point of view, one may regard conformal maps
as an
infinite-dimensional reduction of the dToda hierarchy.

In the rest of this section we  discuss
a  further reduction leading to new solutions
of finite-dimensional WDVV equations.
Consider a class of conformal maps
represented by Laurent polynomials of the form
\beq\label{zpol}
z(w)=rw
+\sum_{l=0}^{N-1}u_l w^{-l}.
\eeq
As proved in~\cite{MWZWZ}, this class of functions
represents conformal maps to domains with a finite number of non-zero
moments, namely, with
$t_{k}=\bar t_k =0$ for $k>N$.
The residue formula~(\ref{reskri})
with $j,k,m \geq 0$, specific to this case reads
\beq\label{resspec}
{\cal F}^{(0)}_{jkm} =
\sum_{a=1}^{N}
\frac{\phi_j(w_a) \phi_k(w_a) \phi_m
(w_a)w_{a}^{N-1}}{P'_N(w_a)Q_N(w_a)}
\eeq
where $P_N(w)=w^N z'(w)$, $Q_N=\bar z'(w^{-1})$
are polynomials of $N$-th degree:
$$
P_N(w)=rw^N -\sum_{k=1}^{N-1}ku_k w^{N-k-1}\,,
\;\;\;\;\;
Q_N(w)=r -\sum_{k=1}^{N-1}k\bar u_k w^{k+1}
$$
and $w_a$ are zeros of $P_N(w)$. They are
inside the unit circle whereas zeros
of $Q_N$ are outside.
Recall that $\phi_j(w)$ for $j\geq 0$
are polynomials in $w$ (of $j$-th degree).
In particular, $\phi_0 (w)=1$.

The algebra
\beq\label{finalg}
\phi_j (w)\phi_k (w) =\sum_{l=0}^{N-1} \check C_{jk}^{l}\phi_l(w)
\;\;\;(\mbox{mod}\, P_N(w))\,,
\;\;\;\;\;j,k=0,1, \ldots , N-1\,
\eeq
is an $N$-dimensional associative algebra
isomorphic to the ring of all polynomials factorized over
the ideal generated by $P_N(w)$.
It is easy to see from the residue formula
(\ref{resspec}) that the structure
constants obey ${\cal F}^{(0)}_{ijk}=\displaystyle{\sum_{l=0}^{N-1}
\check C_{ij}^{l}{\cal F}^{(0)}_{lk0}}$. To do that, one
should apply~(\ref{resspec}) to the $N$-dimensional set
of flows $t_0, t_1, \ldots , t_{N-1}$.
Therefore, we conclude that (logarithm of) the
tau-function for curves~\cite{MWZWZ,KKMWZ}, being restricted
to the space where all the times $t_k$, $\bar t_k$ with
$k>N$ are zero
(and $t_{N}\neq 0$ plays
a role of a parameter), provides a solution to the
finite WDVV equations.
More precisely,
\beq\label{fF}
  \left. \phantom{\frac{a}{b}}
f(t_0, t_1, \ldots , t_{N-1}) =
{\cal F}^{(0)}(t_0; t_1, \ldots , t_{N-1}, t_N, 0,0, \ldots ;
\bar t_1, \ldots , \bar t_N , 0, 0, \ldots )
\right |_{t_N \neq 0, \,\,\,\bar t_j \;\; \mbox{fixed}}
\eeq
solves the WDVV system~(\ref{fullWDVV}) with the matrices
$({\sf F}_{i})_{jk}=\displaystyle{
\frac{\p^3 f}{\p t_i \p t_j \p t_k}}$, $0\leq i,j,k \leq N-1$.
(We stress that the ``antiholomorphic'' times $\bar t_k$
and the highest non-zero time $t_N$ are kept
constant under the differentiation.)

In contrast to solutions to the finite WDVV system discussed
in Sec.\,3.4, the solutions constructed here do not allow one
to switch on the higher flows (in other words there is no
large phase space) since they do not preserve the form~(\ref{zpol}).

For $N=2$ (the curve is an ellipse), the WDVV
are empty since the number
of independent variables is less than three.
The first non-trivial example is
the case $N=3$, where the tau-function is not available
in an explicit form. This makes it difficult to confirm
the result by a direct verification.
One might
find ${\cal F}_{ijk}^{(0)}$ directly from the residue formula
using one or another parameterization of the conformal map
(with enough number of independent parameters).
In this way, we checked that the solutions obtained
do obey WDVV for $N=3$ and $N=4$ using MAPLE\footnote{We thank
A.Mironov for help in the calculation.}.

We expect that a more general class of
finite-dimensional solutions to WDVV is generated by
the tau-function of curves in the case when
$z'(w)$ is a rational function.
We hope to discuss this elsewhere.

\section{Conclusion}

We have shown that given a solution to
the dKP or dToda integrable
hierarchies the Hamiltonians of commuting flows form an
associative (in general infinite dimensional) algebra, isomorphic
either to polynomial ring or to the ring of certain rational functions
(Laurent polynomials).
In the basis provided by integrable hierarchy the structure
constants of the algebra
are expressed through the third order derivatives of the logarithm of
the
tau-function. The algebra becomes finite dimensional in the case of
special solutions known as finite dimensional or algebraic reductions of
the hierarchy.
Such algebras were previously known for the
$N$-KdV reductions of the dKP hierarchy.
We showed that the Hirota's relations in the dispersionless
limit determine the structure constants of the algebra
and the associativity conditions or WDVV equations can be obtained
as a simple consequence of Hirota's equations.

This provides a
motivation to investigate whether any solution
to WDVV equations obeys some
Hirota-type relations for second order derivatives of
the ${\cal F}$.
The theory of the Whitham hierarchy on
Riemann surfaces of arbitrary genus developed in
\cite{KriW} suggests a proper generalization of
the dispersionless Hirota equations and associative algebras to higher
genus.

Starting from the tau-function for conformal maps,
we have constructed a family of solutions to WDVV equations
with finite number of variables. They correspond to some
finite-dimensional associative algebras of a special form.

\section*{Acknowledgements}

We are indebted to H.Braden, E.Getzler, L.Hoevenaars,
A.Mironov, S.Na\-tan\-zon, D.H.Phong, T.Ta\-ke\-be,
and especially to I.Krichever for illuminating discussions.
We are also grateful to L.Hoevenaars and A.Mironov for help
in some MAPLE calculations.
The work by A.B.
was partially supported by CRDF grant UP1-2115,
the work of A.M. was partially supported by
RFBR grant No.~01-01-00539, INTAS
grant No.~99-01782, CRDF grant No.  RP1-2102 (6531) and grant for the
support
of scientific schools No.~00-15-96566, the work of O.R. was 
partially supported by the
DOE ASCI/Alliances Center for
Astrophysical Thermonuclear Flashes at the University of Chicago and
the work of A.Z. was supported in part
by CRDF grant RP1-2102, by grant INTAS-99-0590 and by RFBR grant
01-01-00539.
A.Z and P. W. have been also supported by grants
NSF DMR 9971332 and MRSEC NSF DMR 9808595.

\section*{Appendix}

Here we derive the WDVV equations  for the dToda hierarchy
starting from Hirota's relations~(\ref{eq:50},~\ref{eq:64}).

Let all free
indices in~(\ref{WDVVToda}) be strictly positive. Passing to the
generating functions, we rewrite the equation as
\begin{equation}
      \label{eq:57}
        \frac{1}{w_{12}}
(w_1 D_1 - w_2 D_2) \dd 34 =
\frac{1}{w_{13}}(w_1 D_1 - w_3 D_3 )\dd 24,
\end{equation}
where we set $w^{+}_i =w_i$, $D^{+}_i =D_i$ for simplicity of notation.
Eq.~(\ref{eq:57}) is equivalent to
\begin{equation}
    \label{eq:6}
         w_1(w_{13} \d 134 - w_{12} \d124 - w_{23} \d234 ) = 0,
\end{equation}
which is true due to the identity
\begin{equation}
      \label{eq:58}
      w_{12} \d124 + w_{23} \d234  + w_{31} \d 134 = 0.
\end{equation}
Identity~(\ref{eq:58}) can be easily obtained from~(\ref{eq:51}) by
passing to
the generating functions (multiplying it by $\z4^{-k}$, summing over
$k$) and
then adding two more contributions with cyclic permutations of $1, 2,
3$.  The
same arguments go through without modifications if the index $m$ in
(\ref{WDVVToda}) is non-positive.

Let $j$ be negative, and all others be positive.  Below, we employ the
simplified notation $w^{-}_i =\bar w_i$, $D^{-}_i =\bar D_i$.  In terms
of the
generating functions the WDVV now reads
\begin{equation}
      \label{eq:71}
        \dfrac{(D_1 + \bar D_2 + \pt)\dd34}{w_1\wb2 - 1} =
-\, \frac{(w_1 D_1 -w_3 D_3 )\bar D_2 D_4{\cal F}}{w_1 -w_3}.
\end{equation}
We express $\bar D_2 D_1 D_4 {\cal F}$ and $\bar D_2 D_3 D_4 {\cal F}$
from~(\ref{eq:65}) and substitute them back into~(\ref{eq:71}). It gives
\begin{equation}
      \label{eq:72}
         w_{13}\d134 + w_1\dd14_0 - w_3\dd34_0 = 0
\end{equation}
in which one readily recognizes eq.~(\ref{eq:51}), which concludes the
proof
for this case.

At last, let $j,k$ be negative, and $i,m$ be positive.  In this case we
have
the following WDVV to prove:
\begin{equation}
      \label{eq:74}
         \frac{(
\bar D_3 D_4 {\cal F}_0 +
\bar D_3 D_1 D_4 {\cal F} +
\bar D_3 \bar D_2 D_4 {\cal F})}{(w_1\wb2 - 1)} =
         \frac{(
\bar D_2 D_4 {\cal F}_0 +
\bar D_2 D_1 D_4 {\cal F} +
\bar D_2 \bar D_3 D_4 {\cal F}
)}{(w_1\wb3 - 1)}.
\end{equation}
As in a previous case use analog of eq.~(\ref{eq:65}) to substitute for
$\bar
D_3 D_1 D_4$ and $\bar D_2 D_1 D_4$. Also note that $\bar D_2 \bar D_3
D_4$
together with $\bar D_3 D_4 \pt{\cal F}$ and $\bar D_2 D_4 \pt{\cal F}$
form
analog of~(\ref{eq:72}). All these facts combined together show
that~(\ref{eq:74}) holds.
which concludes the proof of WDVV for the Toda case.

\end{document}